\documentclass{article}

\usepackage{bm}
\usepackage{amsmath}
\usepackage{amssymb}

\usepackage[onehalfspacing]{setspace}
\usepackage[utf8]{inputenc}
\usepackage[T1]{fontenc}
\usepackage[hidelinks]{hyperref}
\usepackage{graphicx}
\usepackage{tikz}
\usepackage{algorithm}
\usepackage{algorithmic}
\usepackage{csquotes}
\usepackage{geometry}
\usepackage[sorting=none]{biblatex}
\usepackage{microtype}
\usepackage{cleveref}
\usepackage{newtxtext,newtxmath}
\newcommand{\mathbfit}[1]{\bm{#1}}
\newcommand{\rmd}{\mathrm{d}}
\newcommand{\vGL}{\mathbfit{g}}
\newcommand{\vGG}{\mathbfit{G}}
\newcommand{\iGL}{\mathbfit{g}_i}
\newcommand{\iGG}{\mathbfit{G}_i}
\newcommand{\vHL}{\mathbfit{h}}
\newcommand{\vHG}{\mathbfit{H}}
\newcommand{\rHL}{\mathbfit{h}_r}
\newcommand{\cHL}{\mathbfit{h}_c}
\newcommand{\rHG}{\mathbfit{H}_r}
\newcommand{\cHG}{\mathbfit{H}_c}
\newcommand{\aN}{\mathcal{A}}

\usepackage{authblk}

\title{Vectorized Sparse Second-Order Forward Automatic Differentiation for Optimal Control Direct Methods}
\author{Yilin Zou\thanks{Ph.D. candidate, School of Aerospace Engineering, zouyl22@mails.tsinghua.edu.cn.} }
\author{Fanghua Jiang\thanks{Associate Professor, School of Aerospace Engineering, jiangfh@tsinghua.edu.cn (Corresponding Author).}}
\affil{Tsinghua University, Beijing, China}
\date{\today}

\begin{document}

\maketitle

\begin{abstract}
    Direct collocation methods are widely used numerical techniques for solving optimal control problems. The discretization of continuous-time optimal control problems transforms them into large-scale nonlinear programming problems, which require efficient computation of first- and second-order derivatives. To achieve computational efficiency, these derivatives must be computed in sparse and vectorized form, exploiting the problem's inherent sparsity structure.
    This paper presents a vectorized sparse second-order forward automatic differentiation framework designed for direct collocation methods in optimal control. The method exploits the problem's sparse structure to efficiently compute derivatives across multiple mesh points. By incorporating both scalar and vector nodes within the expression graph, the approach enables effective parallelization and optimized memory access patterns while maintaining flexibility for complex problems. The methodology is demonstrated through application to a prototype optimal control problem. A complete implementation for multi-phase optimal control problems is available as an open-source package, supporting both theoretical research and practical applications.
\end{abstract}

\section{Introduction}
Optimal control theory is fundamental to many scientific and engineering applications. The goal of optimal control is to determine control inputs as functions over time that minimize (or maximize) an objective function subject to system dynamics and constraints. Most optimal control problems cannot be solved analytically, requiring numerical methods for their solution.

Numerical solution methods for optimal control problems can be broadly categorized into indirect methods and direct methods~\parencite{bettsSurveyNumericalMethods1998}. Indirect methods, based on calculus of variations and Pontryagin's maximum principle, derive first-order necessary conditions for optimality. These conditions lead to a two-point boundary value problem (TPBVP) involving both state and costate variables. The TPBVP is typically solved using shooting methods, where initial guesses for the costate variables are iteratively adjusted until convergence is achieved. While indirect methods can provide accurate solutions and theoretical insights, they present several practical challenges: the costate variables lack physical interpretation, making it difficult to provide good initial guesses; the resulting TPBVP is often highly sensitive to initial conditions, leading to convergence difficulties; and deriving the necessary conditions can be complex for problems with path or algebraic constraints~\parencite{bettsPracticalMethodsOptimal2010}.

Direct methods discretize the continuous optimal control problem into a finite-dimensional nonlinear programming (NLP) problem. Consider the Legendre-Gauss-Radau (LGR) pseudospectral method as an example~\parencite{gargDirectTrajectoryOptimization2011a,gargPseudospectralMethodsSolving2011,gargUnifiedFrameworkNumerical2010a}. The state and control variables are approximated by piecewise interpolation polynomials at LGR quadrature points. The objective, constraint, and dynamics functions are integrated or enforced at these nodes, transforming the problem into an NLP problem~\parencite{gargDirectTrajectoryOptimization2011}. The resulting NLP problem has a large number of variables and constraints with sparse structure. Therefore, it is essential to compute derivatives of the objective and constraint functions of the discretized problem in sparse and vectorized form for the NLP solver. The NLP solver must handle these sparse derivatives efficiently and implement robust optimization algorithms for large-scale problems. Popular NLP solvers include IPOPT~\parencite{wachterImplementationInteriorpointFilter2006}, SNOPT~\parencite{gillSNOPTSQPAlgorithm2005}, and KNITRO~\parencite{byrdKnitroIntegratedPackage2006}, which use interior-point methods or sequential quadratic programming algorithms to solve large-scale NLP problems. Direct methods offer greater flexibility for handling complex objectives and constraints and are more robust to initial guess selection, making them popular choices for solving optimal control problems in practice.

To compute the derivatives of the objective and constraint functions efficiently for the discretized problem, one common approach is to use finite difference methods. For example, GPOPS-II~\parencite{pattersonGPOPSIIMATLABSoftware2014} is a MATLAB software package for solving multiple-phase optimal control problems using pseudospectral methods. It employs sparse finite difference methods to compute derivatives by perturbing each variable and exploiting the known sparsity structure. The software includes optimizations such as determining the minimum perturbation size to achieve reasonable accuracy and efficiently grouping variables to minimize function evaluations.

However, finite difference methods have several limitations: they can reduce optimization convergence rates, suffer from truncation and roundoff errors, and may produce inaccurate results for ill-conditioned problems. Given these challenges in accurately computing derivatives using finite differences, automatic differentiation (AD) emerges as a powerful alternative. AD systematically applies the chain rule to compute derivatives by tracking the sequence of elementary operations that comprise a function evaluation. Unlike finite differences, AD computes derivatives with machine precision without suffering from truncation or roundoff errors.

Several software packages have been developed that use AD to compute derivatives for optimal control direct methods. PSOPT~\parencite{becerraSolvingComplexOptimal2010} is a software package for optimal control and nonlinear optimization problems that uses ADOL-C~\parencite{griewankAlgorithm755ADOLC1996} for automatic differentiation. ADOL-C employs operator overloading in C++ to record arithmetic operations and compute derivatives in either forward or reverse mode. Tiger Optimization Software (TOPS)~\parencite{sowellIntroducingTigerOptimization2024} utilizes the ADiGator~\parencite{weinsteinAlgorithm984ADiGator2017} automatic differentiation tool to compute derivatives for large-scale optimization problems. These software packages provide efficient and accurate derivatives for optimal control problems, enabling the use of modern optimization algorithms that require first- and second-order derivatives. In this paper, we develop a vectorized sparse second-order automatic differentiation method specifically designed for optimal control problems discretized with direct methods. This method is tailored to the structure of optimal control problems, providing benefits in terms of computational transparency, fully vectorized operations, and efficient handling of second-order derivatives.

AD has two main variants: forward mode and reverse mode. Forward mode AD propagates derivatives from inputs to outputs through the computation graph, making it efficient when there are few inputs and many outputs. Reverse mode AD propagates derivatives backward from outputs to inputs, making it efficient when there are many inputs and few outputs~\parencite{margossianReviewAutomaticDifferentiation2019}. For optimal control problems discretized with direct methods, each mesh point has $n_x + n_u + 2$ input variables, where $n_x$ is the number of state variables, $n_u$ is the number of control variables, and the additional 2 variables represent the initial and final time. The number of outputs is $n_x + n_c + 1$, where $n_c \geq 0$ is the number of path constraints, and $n_x + 1$ represents the dynamics equations and the objective function. Since $n_u$ is typically small in practice, the number of inputs is generally comparable to the number of outputs. Therefore, we choose forward mode AD because its propagation rule is simpler and more straightforward to implement, particularly for second-order derivatives.

This paper develops a vectorized sparse second-order forward automatic differentiation method that efficiently computes gradients and Hessians of objective and constraint functions in direct collocation methods. This approach integrates well with modern optimization algorithms that require sparse second-order derivatives in coordinate list (COO) format. The method is designed for efficiency by exploiting the sparse and vectorized structure of the problem. It has $O(N)$ space and time complexity, where $N$ is the number of mesh points, which aligns with the problem's inherent sparsity structure. The implementation is fully vectorized across mesh points, enabling efficient execution on modern hardware with Single Instruction, Multiple Data (SIMD) capabilities. Consequently, the method achieves high convergence rates and numerical stability while maintaining computational efficiency.

We demonstrate the methodology using a prototype optimal control problem to provide clear exposition. While this example is relatively simple, the methodology is general and can handle more complex multi-phase optimal control problems. The implementation is available in our open-source software package pockit\footnote{The source code is available at \urlstyle{same}\url{https://github.com/zouyilin2000/pockit}} (Python Optimal Control KIT). The library solves multi-phase optimal control problems using both Legendre-Gauss-Radau and Legendre-Gauss-Lobatto collocation schemes. The software provides both a framework for theoretical studies of direct collocation methods and practical solutions for optimal control problems across diverse domains.

The paper is organized as follows: \Cref{sec:problem} introduces the optimal control problem formulation and its discretization using the LGR collocation method. \Cref{sec:ad} presents the forward algorithm for computing gradients and Hessians using a non-vectorized expression graph. \Cref{sec:vectorized_ad} extends the method to vectorized expression graphs, enabling efficient computation across multiple mesh points. \Cref{sec:graph_structure} demonstrates the application of the vectorized sparse second-order forward automatic differentiation method to the prototype optimal control problem. \Cref{sec:conclusion} concludes the paper and discusses future research directions.

Throughout this article, we assume that matrices are stored in column-major order in memory. The algorithm works equally well with row-major systems but requires appropriate transposition to maintain cache locality and computational efficiency. 

\section{Problem Formulation and Discretization}\label{sec:problem}
\subsection{Prototype Problem Formulation}
To illustrate the automatic differentiation methodology clearly, we consider a single-phase optimal control problem without path constraints. The continuous-time problem is formulated as:
\begin{align}
    \min_{\mathbfit{x}, \mathbfit{u}, t_0, t_f} & \quad J = \int_{t_0}^{t_f} f(\mathbfit{x}, \mathbfit{u}, t)\, \rmd t \\
    \text{s.t.} & \quad \dot{\mathbfit{x}} = \mathbfit{g}(\mathbfit{x}, \mathbfit{u}, t)\\
    & \quad \text{Boundary conditions on $\mathbfit{x}(t_0)$, $\mathbfit{x}(t_f)$, $t_0$, and $t_f$}
\end{align}
where $\mathbfit{x} \in \mathbb{R}^{n_x}$ represents the state variables, $\mathbfit{u} \in \mathbb{R}^{n_u}$ represents the control variables, $t_0$ and $t_f$ are the initial and final times, $f$ is the objective function, and $\mathbfit{g}$ describes the system dynamics. The boundary conditions may specify fixed or free values for each state variable and the time endpoints.

\subsection{LGR Collocation}
We employ the LGR collocation method for problem discretization. The LGR method is a widely used direct collocation approach, though the developed methodology is general and can be applied to other collocation schemes. In the LGR method, the collocation nodes are the roots of a superposition of Legendre polynomials. The state and control variables are approximated using polynomial interpolation between collocation points, the objective function is integrated using Gaussian quadrature, and the dynamics are enforced at each collocation point. For a comprehensive treatment of the LGR collocation method, we refer readers to the literature~\parencite{gargDirectTrajectoryOptimization2011a,gargPseudospectralMethodsSolving2011,gargUnifiedFrameworkNumerical2010a}. 

We use $N$ collocation nodes to approximate each state and control variable over the time domain. These nodes, scaled to the interval $[0, 1]$, are denoted as the mesh points $\mathbfit{M}$, which form a column vector of length $N$. The time at the mesh points is given by $\mathbfit{T} = t_0 + \mathbfit{M} (t_f - t_0) = \left(1 - \mathbfit{M}\right) t_0 + \mathbfit{M} t_f$.

Let $\mathbfit{X}$ and $\mathbfit{U}$ denote the state and control variables at the mesh points, representing $N \times n_x$ and $N \times n_u$ matrices, respectively. Let $\mathbfit{F}$ and $\mathbfit{G}$ denote the objective and constraint functions evaluated at these collocation points, with dimensions $N \times 1$ and $N \times n_x$ respectively. The values of $\mathbfit{F}$ and $\mathbfit{G}$ at each mesh point are determined by evaluating the objective function $f$ and the dynamics function $\mathbfit{g}$ at the corresponding rows of $\mathbfit{X}$, $\mathbfit{U}$, and $\mathbfit{T}$.

To serve as non-collocation nodes at the last subinterval and to ensure that the boundary conditions are satisfied, we introduce an additional non-collocation node at $t_f$ for each state variable. The state variables at all $N + 1$ nodes are denoted by $\overline{\mathbfit{X}}$, an $(N + 1) \times n_x$ matrix whose first $N$ rows correspond to $\mathbfit{X}$. The initial and final rows of $\overline{\mathbfit{X}}$, denoted as $\mathbfit{X}_0$ and $\mathbfit{X}_f$ respectively, are used to enforce the boundary conditions in the discretized problem.

Let $\mathbfit{W}$ denote the column vector of quadrature weights of length $N$, and $\mathbfit{D}$ represent the $N \times (N + 1)$ LGR derivative matrix. The quadrature weights enable numerical integration of the objective function, while the derivative matrix computes state variable derivatives at the collocation points. With these definitions, the continuous-time optimal control problem is discretized as:
\begin{align}
    \min_{\overline{\mathbfit{X}}, \mathbfit{U}, t_0, t_f} & \quad J = \mathbfit{W}^\mathrm{T} \mathbfit{F}\Delta t \label{eq:discrete_obj}\\
    \text{s.t.} & \quad \mathbfit{D} \overline{\mathbfit{X}} = \mathbfit{G} \Delta t \label{eq:discrete_dyn}\\
    & \quad \text{Boundary conditions on $\mathbfit{X}_0$, $\mathbfit{X}_f$, $t_0$, and $t_f$}\label{eq:discrete_bc}
\end{align}
where $\Delta t = t_f - t_0$ represents the time interval length. The discretized problem is solved using NLP solvers for large-scale optimization, which require gradients and Hessians of the objective and constraint functions with respect to the optimization variables. Since the number of optimization variables is large, the derivatives must be computed and provided to the NLP solver in sparse format for computational efficiency.

\subsection{Analysis of the Problem Structure}
A preliminary analysis of the structure of the discretized problem identifies the computational challenges in derivative computation. The boundary conditions in \cref{eq:discrete_bc} involving $\mathbfit{X}_0$, $\mathbfit{X}_f$, $t_0$, and $t_f$ can be handled directly since these variables appear explicitly in the discretized formulation. The left-hand side of the dynamic constraints in \cref{eq:discrete_dyn} is linear with respect to $\overline{\mathbfit{X}}$, so its gradient corresponds to the non-zero elements of the derivative matrix $\mathbfit{D}$, and its Hessian is zero. Similarly, the quadrature weights $\mathbfit{W}$ in \cref{eq:discrete_obj} are constant and represent a linear transformation of the $\mathbfit{F} \Delta t$ vector. They can be handled by multiplying the gradient and Hessian of $\mathbfit{F} \Delta t$ by the corresponding entries of $\mathbfit{W}$. The primary computational challenge lies in computing the gradient and Hessian of the terms $\mathbfit{F}\Delta t$ and $\mathbfit{G}\Delta t$, which share the same structure. In the following sections, we focus on developing the methodology for computing the gradient and Hessian of $\mathbfit{F}\Delta t$; this approach can then be applied column-wise to $\mathbfit{G}\Delta t$ using the same procedure.

\section{Sparse Forward Automatic Differentiation}\label{sec:ad}
To efficiently compute the gradient and Hessian of the objective and constraint functions, we introduce a forward automatic differentiation framework that systematically captures relationships between input, intermediate, and output variables. Since these variables have no circular dependencies, we can represent their relationships as a directed acyclic graph (DAG), called an expression graph. In this structure, each node represents a variable, with edges indicating functional dependencies between variables.

In this section, we present the theoretical foundation of our non-vectorized expression graph data structure and develop the forward algorithm for gradient and Hessian computation.

\subsection{Sparse Expression Graph}
The expression graph comprises nodes representing input, intermediate, and output variables. Each node contains the following components:
\begin{itemize}
    \item Reference to argument nodes $\aN$:
        This set contains references to variables on which the node depends. The partial derivatives with respect to these arguments are assumed to be analytically or numerically computed.
        \item Partial gradient specification via index vector $\iGL$ and value vector $\vGL$:
        These row vectors have equal length $n_g \leq \|\aN\|$. The corresponding entries of $\left(\iGL, \vGL\right)$ form pairs $\left(g_i, g\right)$, where $g_i$ indicates the argument index and $g$ represents the corresponding partial derivative value. This sparse representation efficiently captures non-zero partial derivatives.
        \item Full gradient specification via index vector $\iGG$ and value vector $\vGG$:
        Similar to the partial gradient, these row vectors have equal length, with corresponding entries denoting the index and value of the full gradient with respect to input variables.
        These components are initially defined only for input variables. For intermediate and output variables, these values are computed through the forward algorithm.
        \item Partial Hessian specification via index vectors $\rHL, \cHL$ and value vector $\vHL$:
        These row vectors, of equal length $n_h \leq \|\aN\| \left(\|\aN\| + 1\right) / 2$, 
        represent half of the symmetric Hessian matrix with respect to the arguments in COO format. The indices $\rHL$ and $\cHL$ represent the row and column indices of the non-zero elements, while $\vHL$ contains the corresponding values. 
        To reduce computational cost and storage requirements, only half of the Hessian matrix is stored: For off-diagonal elements, either the upper or lower triangular element is stored; for diagonal elements, the element is stored with half its actual partial Hessian value. This format enables efficient implementation of the forward algorithm.
        \item Full Hessian specification via index vectors $\rHG, \cHG$ and value vector $\vHG$:
        These row vectors of equal length store the COO format of half of the full Hessian matrix with respect to input variables. The half-Hessian follows the same convention as the partial Hessian.
        Similar to the full gradient, these components are initially defined only for input variables. The forward algorithm computes the full Hessian for intermediate and output variables.
\end{itemize}

\subsection{Forward Algorithm}
We now present the forward algorithm for computing gradients and Hessians of a node, given its partial derivatives and the full derivatives of its arguments. The acyclic structure of the graph ensures that a topological ordering exists, which allows complete derivative computation in a single forward pass. The algorithm consists of two phases: computing gradients and computing Hessians.

\subsubsection{Forward Gradient}
Consider a node $N$ with arguments $\aN = \left\lbrace A^{(1)}, A^{(2)}, \ldots\right\rbrace$. The index and value vectors of $N$'s full gradient are computed as:
\begin{align}
    N_{\iGG} &= \text{concat}\left(A^{(g_i)}_{\iGG} : g_i \in \iGL\right)\label{eq:fg_index}\\
    N_{\vGG} &= \text{concat}\left(g A^{(g_i)}_{\vGG} : \left(g_i, g\right) \in \left(\iGL, \vGL\right)\right)\label{eq:fg_value}
\end{align}
where the notation in brackets uses list comprehension and $\text{concat}$ denotes vector concatenation. The iterations in Equations \eqref{eq:fg_index} and \eqref{eq:fg_value} process each pair of partial indices and values, multiplying the partial gradient by the corresponding argument's full gradient and concatenating the results. The algorithm's correctness follows directly from the chain rule of differentiation:
\begin{equation}
    \frac{\partial N}{\partial x} = \sum_{g_i\in\iGL} \frac{\partial N}{\partial A^{(g_i)}} \frac{\partial A^{(g_i)}}{\partial x} = \sum_{\left(g_i, g\right) \in \left(\iGL, \vGL\right)} g \frac{\partial A^{(g_i)}}{\partial x}
\end{equation}

\subsubsection{Forward Hessian}
Consider a node $N$ with arguments $\aN = \left\lbrace A^{(1)}, A^{(2)}, \ldots\right\rbrace$.
The full Hessian comprises two components: the product of the node's partial gradients with the arguments' full Hessians, and the product of the node's partial Hessian with the arguments' full gradients.

The first component is computed as:
\begin{align}
    N_{\rHG}^{\text{gh}} &= \text{concat}\left(A^{(g_i)}_{\rHG} : g_i \in \iGL\right)\label{eq:fh_gh_r}\\
    N_{\cHG}^{\text{gh}} &= \text{concat}\left(A^{(g_i)}_{\cHG} : g_i \in \iGL\right)\label{eq:fh_gh_c}\\
    N_{\vHG}^{\text{gh}} &= \text{concat}\left(g A^{(g_i)}_{\vHG} : \left(g_i, g\right) \in \left(\iGL, \vGL\right)\right)\label{eq:fh_gh_v}
\end{align}
These equations compute the product of the node's partial gradients with the full Hessians of its arguments. The corresponding indices and values are concatenated to form the first component of the node's full Hessian.

The second component is computed as:
\begin{align}
    \left(N_{\rHG}^{\text{hg}}, N_{\cHG}^{\text{hg}}\right) &= \text{concat}\left(A^{(h_r)}_{\iGG} \times A^{(h_c)}_{\iGG}: \left(h_r, h_c\right) \in \left(\rHL, \cHL\right)\right)\label{eq:fh_hf_rc}\\
    N_{\vHG}^{\text{hg}} &= \text{concat}\left(h  \left(A^{(h_r)}_{\vGG} \otimes A^{(h_c)}_{\vGG}\right): \left(h_r, h_c, h\right) \in \left(\rHL, \cHL, \vHL\right)\right)\label{eq:fh_hf_v}
\end{align}
These equations compute the second component of the full Hessian through Cartesian and Kronecker products of the arguments' full gradients. Each element in the node's partial Hessian corresponds to a pair of argument nodes $A^{(h_r)}$ and $A^{(h_c)}$ with partial Hessian value $h$. The full gradients of these argument nodes are combined through a Cartesian product and multiplied by the partial Hessian value to obtain the corresponding full Hessian contribution.
In \cref{eq:fh_hf_rc}, the row and column indices are obtained from the Cartesian product of the argument full gradient indices. The first elements of the resulting pairs form the row indices, while the second elements form the column indices. In \cref{eq:fh_hf_v}, the values are computed using the Kronecker product, where each element of the first argument node's full gradient is multiplied with every element of the second argument node's full gradient. The resulting vector is then scaled by the partial Hessian value to obtain the final full Hessian values.

The complete full Hessian of node $N$ is obtained by concatenating the two components:
\begin{align}
    N_{\rHG} &= \text{concat}\left(N_{\rHG}^{\text{gh}}, N_{\rHG}^{\text{hg}}\right)\label{eq:fh_r}\\
    N_{\cHG} &= \text{concat}\left(N_{\cHG}^{\text{gh}}, N_{\cHG}^{\text{hg}}\right)\label{eq:fh_c}\\
    N_{\vHG} &= \text{concat}\left(N_{\vHG}^{\text{gh}}, N_{\vHG}^{\text{hg}}\right)\label{eq:fh_v}
\end{align}

The algorithm's correctness follows from the chain rule of differentiation. Since $\rHL, \cHL, \vHL$ represent half of the symmetric Hessian matrix, the second partial derivative of $N$ with respect to input variables $x_i, x_j$ is:
\begin{align}
    \frac{\partial^2 N}{\partial x_i \partial x_j} 
    = \sum_{\left(g_i, g\right) \in \left(\iGL, \vGL\right)} g \frac{\partial^2 A^{(g_i)}}{\partial x_i \partial x_j}
    + \sum_{\left(h_r, h_c, h\right) \in \left(\rHL, \cHL, \vHL\right)} h \left(\frac{\partial A^{(h_r)}}{\partial x_i} \frac{\partial A^{(h_c)}}{\partial x_j} + \frac{\partial A^{(h_r)}}{\partial x_j} \frac{\partial A^{(h_c)}}{\partial x_i}\right)
\end{align}
The first term represents the contribution from the node's partial gradient multiplied by the arguments' Hessians, while the second term captures the contribution from the node's partial Hessian multiplied by the arguments' gradients. For off-diagonal elements ($i \neq j$), the Cartesian product produces both required cross terms. For diagonal elements ($i = j$), the Cartesian product yields a single term, which matches our half-Hessian storage convention where diagonal elements are stored with half their actual value.

By applying the forward algorithm to each node in the expression graph in topological order, we can efficiently compute the full gradient and Hessian of output nodes corresponding to the objective and constraint functions in the optimal control problem. The forward algorithms for computing the full gradient and Hessian of a node are summarized in \cref{alg:forward_gradient,alg:forward_hessian}, respectively.

\begin{algorithm}
    \caption{Forward Algorithm for Full Gradient Computation}
    \label{alg:forward_gradient}
    \begin{algorithmic}
        \STATE \textbf{Input:} Node $N$ with arguments $\aN = \left\lbrace A^{(1)}, A^{(2)}, \ldots\right\rbrace$, partial gradients with index $\iGL$ and value $\vGL$
        \STATE \textbf{Output:} Full gradient of $N$ with index $\iGG$ and value $\vGG$
        \STATE \textbf{Initialize:} $\iGG \gets [], \vGG \gets []$
        \FOR{$\left(g_i, g\right) \in \text{zip}\left(\iGL, \vGL\right)$}
            \STATE extend $\iGG$ with $A^{(g_i)}_{\iGG}$
            \STATE extend $\vGG$ with $g A^{(g_i)}_{\vGG}$
        \ENDFOR
    \end{algorithmic}
\end{algorithm}

\begin{algorithm}
    \caption{Forward Algorithm for Full Hessian Computation}
    \label{alg:forward_hessian}
    \begin{algorithmic}
        \STATE \textbf{Input:} Node $N$ with arguments $\aN = \left\lbrace A^{(1)}, A^{(2)}, \ldots\right\rbrace$, partial gradients with index $\iGL$ and value $\vGL$, partial Hessian with row index $\rHL$, column index $\cHL$, and value $\vHL$
        \STATE \textbf{Output:} Full Hessian of $N$ with row index $\rHG$, column index $\cHG$, and value $\vHG$
        \STATE \textbf{Initialize:} $\rHG^{\text{gh}} \gets [], \cHG^{\text{gh}} \gets [], \vHG^{\text{gh}} \gets [], \rHG^{\text{hg}} \gets [], \cHG^{\text{hg}} \gets [], \vHG^{\text{hg}} \gets []$
        \FOR{$\left(g_i, g\right) \in \text{zip}\left(\iGL, \vGL\right)$}
            \STATE extend $\rHG^{\text{gh}}$ with $A^{(g_i)}_{\rHG}$
            \STATE extend $\cHG^{\text{gh}}$ with $A^{(g_i)}_{\cHG}$
            \STATE extend $\vHG^{\text{gh}}$ with $g A^{(g_i)}_{\vHG}$
        \ENDFOR
        \FOR{$\left(h_r, h_c, h\right) \in \text{zip}\left(\rHL, \cHL, \vHL\right)$}
            \STATE extend $\rHG^{\text{hg}}$ with first element of $A^{(h_r)}_{\iGG} \times A^{(h_c)}_{\iGG}$
            \STATE extend $\cHG^{\text{hg}}$ with second element of $A^{(h_r)}_{\iGG} \times A^{(h_c)}_{\iGG}$
            \STATE extend $\vHG^{\text{hg}}$ with $h \left(A^{(h_r)}_{\vGG} \otimes A^{(h_c)}_{\vGG}\right)$
        \ENDFOR
        \STATE $\rHG \gets \text{concat}\left(\rHG^{\text{gh}}, \rHG^{\text{hg}}\right)$
        \STATE $\cHG \gets \text{concat}\left(\cHG^{\text{gh}}, \cHG^{\text{hg}}\right)$
        \STATE $\vHG \gets \text{concat}\left(\vHG^{\text{gh}}, \vHG^{\text{hg}}\right)$
    \end{algorithmic}
\end{algorithm}

\section{Vectorized Sparse Automatic Differentiation}\label{sec:vectorized_ad}
The non-vectorized sparse AD framework provides a foundation for computing gradients and Hessians. However, discretized optimal control problems involve variables that scale with the number of mesh points $N$, which can be large in practical applications. A key observation is that computations of the objective and constraint functions $\mathbfit{F}\Delta t$ and $\mathbfit{G}\Delta t$ at different mesh points are independent. To achieve efficient performance for large-scale problems, we extend the AD framework to handle vectorized computations across mesh points. This extension exploits the independent nature of computations at different mesh points, enabling parallelization and SIMD execution on modern processors while improving memory access patterns.

\subsection{Data Structure of the Vectorized Expression Graph}
The vectorized expression graph distinguishes between two types of nodes: scalar nodes and vector nodes. 
Scalar nodes represent scalar variables such as $t_0$, $t_f$, and $\Delta t$, each with a single value. Vector nodes represent variables at mesh points such as $\mathbfit{X}_i$, $\mathbfit{U}_i$, and $\mathbfit{F}$, where $\mathbfit{X}_i$ and $\mathbfit{U}_i$ denote the $i$-th column of $\mathbfit{X}$ and $\mathbfit{U}$, respectively. All vector nodes have uniform length $N$. Input variables $\mathbfit{X}_i$ and $\mathbfit{U}_i$ are stored contiguously in the discretized problem layout, allowing us to store only the leading index of vector nodes and optimize memory access patterns.

The data structure of scalar nodes remains identical to the non-vectorized framework.
For vector nodes, the index vectors $\iGL$, $\iGG$, $\rHL$, $\cHL$, $\rHG$, and $\cHG$ remain row vectors, storing only the leading index as described above. The value vectors $\vGL$, $\vGG$, $\vHL$, and $\vHG$ become matrices with $N$ rows, where each row corresponds to a mesh point.
The vectorized expression graph enforces the following constraint: scalar nodes may depend only on other scalar nodes, while vector nodes may depend on both scalar and vector nodes.

\subsection{Forward Algorithm for Vectorized Gradient and Hessian}
The vectorized forward algorithm maintains the fundamental structure of its non-vectorized counterpart, with operations performed on matrix columns rather than scalar elements. 
A key extension is the treatment of dependencies between vector and scalar nodes: when a vector node depends on a scalar node, the scalar node's derivatives are automatically extended to vector form by replicating the row vectors $\vGL$, $\vGG$, $\vHL$, and $\vHG$ across $N$ rows to form matrices.
This transformation can be efficiently implemented through broadcasting operations rather than explicit memory allocation. The correctness of this approach follows directly from the chain rule applied to each mesh point independently.

Since the forward algorithms (\cref{alg:forward_gradient,alg:forward_hessian}) contain no conditional branching, the vectorized implementation can utilize optimized matrix-vector and matrix-matrix operations available in modern linear algebra libraries. Note that the Kronecker product in \cref{eq:fh_hf_v} operates only along rows to produce correct results.

\section{Graph Structure of the Prototype Problem}\label{sec:graph_structure}
In this section, we construct and analyze the vectorized expression graph for the optimal control problem presented in \Cref{sec:problem}. As established previously, the objective and constraint functions $\mathbfit{F}\Delta t$ and $\mathbfit{G}\Delta t$ share the same structural foundation. We focus our discussion on the graph for $\mathbfit{F}\Delta t$, noting that the methodology for $\mathbfit{G}\Delta t$ follows by applying the same procedure column-wise.

\begin{figure}
    \centering
    \definecolor{scalar color}{HTML}{8C2336}
    \definecolor{vector color}{HTML}{022e54}
    \begin{tikzpicture}[
        scalar/.style={circle, draw=scalar color, very thick, inner sep=0pt, minimum width=.9cm},
        vector/.style={circle, draw=vector color, very thick, inner sep=0pt, minimum width=.9cm},
        every edge/.style={->, very thick},
    ]
        \def \xw {1.6}
        \def \yw {1.9}
        \node [scalar] (t0) at (0, 0) {$t_0$};
        \node [scalar] (tf) at (\xw, 0) {$t_f$};
        \node [vector] (X1) at (2*\xw, 0) {$\mathbfit{X}_1$};
        \node (Xdot) at (2.5*\xw, 0) {$\cdots$};
        \node [vector] (XN) at (3*\xw, 0) {$\mathbfit{X}_{n_x}$};
        \node [vector] (U1) at (4*\xw, 0) {$\mathbfit{U}_1$};
        \node (Udot) at (4.5*\xw, 0) {$\cdots$};
        \node [vector] (UN) at (5*\xw, 0) {$\mathbfit{U}_{n_u}$};

        \node [scalar] (dt) at (\xw, -\yw) {$\Delta t$};
        \node [vector] (T) at (2*\xw, -\yw) {$\mathbfit{T}$};
        \node [vector] (F) at (3*\xw, -\yw) {$\mathbfit{F}$};
        \node [vector] (Fdt) at (4*\xw, -\yw) {$\mathbfit{F}\Delta t$};

        \draw[->] (t0.south east) -- (dt.north west);
        \draw[->] (tf.south) -- (dt.north);
        \draw[->] (t0.south east) -- (T.north west);
        \draw[->] (tf.south east) -- (T.north west);
        \draw[->] (X1.south east) -- (F.north west);
        \draw[->] (XN.south) -- (F.north);
        \draw[->] (U1.south west) -- (F.north east);
        \draw[->] (UN.south west) -- (F.north east);
        \draw[->] (T.east) -- (F.west);
        \draw[->] (F.east) -- (Fdt.west);
        \draw[->] (dt.south east) to [out=-40, in=220] (Fdt.south west);
    \end{tikzpicture}
    \caption{Vectorized expression graph for $\mathbfit{F}\Delta t$. Scalar nodes are shown in red and vector nodes in blue. Directed edges indicate functional dependencies between variables.}
    \label{fig:graph}
\end{figure}
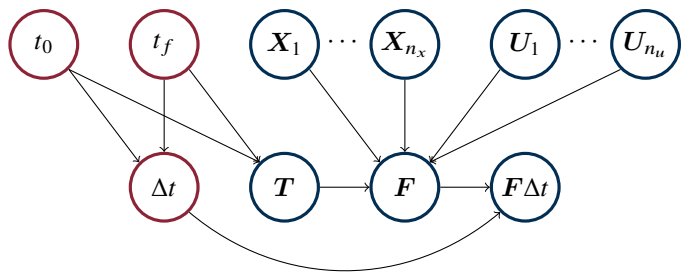

The structure of the expression graph is shown in \Cref{fig:graph}. The first row contains the input variables $t_0$, $t_f$, $\mathbfit{X}_1$, $\ldots$, $\mathbfit{X}_{n_x}$, $\mathbfit{U}_1$, $\ldots$, $\mathbfit{U}_{n_u}$. The second row contains the intermediate and output variables $\Delta t$, $\mathbfit{T}$, $\mathbfit{F}$, and $\mathbfit{F}\Delta t$. Directed edges between nodes indicate functional dependencies, with the direction showing how information propagates through the computational graph. Scalar nodes are shown in red, while vector nodes are shown in blue.

The full gradients and Hessians for input nodes, and partial gradients and Hessians for intermediate and output nodes, are specified as follows:
\begin{itemize}
    \item For input nodes, the full gradient indices correspond to the ordering of input variables, with values set to unity. The full Hessian indices and values are initialized as empty vectors.
    \item For the intermediate node $\Delta t$, the partial gradient with respect to $t_0$ and $t_f$ is $[-1, 1]$. For $\mathbfit{T}$, the partial gradient with respect to these variables is $\left[1-\mathbfit{M}, \mathbfit{M}\right]$. Both nodes have zero partial Hessians.
    \item The partial derivatives of $\mathbfit{F}$ are determined by the objective function $f$. Our implementation in pockit uses symbolic differentiation of user-provided objective functions to compute these derivatives, which are then compiled into efficient vectorized functions for evaluation at $\left(\mathbfit{X}, \mathbfit{U}, \mathbfit{T}\right)$. Alternatively, AD methods can be applied to compute these derivatives directly from the user-provided functions.
    \item The node $\mathbfit{F}\Delta t$ has partial gradient $\left[\mathbfit{F}, \mathbfit{1}_{N \times 1}\right]$ with respect to its arguments $\mathbfit{F}$ and $\Delta t$. Its partial Hessian consists of unit values for the off-diagonal terms.
\end{itemize}

With these derivative specifications, application of the forward algorithm yields the full gradient and Hessian of $\mathbfit{F}\Delta t$. After element-wise multiplication by the quadrature weights $\mathbfit{W}$, the results are directly available in COO format for the NLP solver. The same procedure is applied column-wise to compute the derivatives of $\mathbfit{G}\Delta t$, where each column forms a separate vector node connected to the expression graph in the same manner as $\mathbfit{F}\Delta t$.

\section{Conclusion}\label{sec:conclusion}
Direct collocation methods provide a powerful framework for solving complex optimal control problems, but they require efficient computation of gradients and Hessians for modern optimization algorithms. To achieve high computational efficiency, these derivatives must be computed in sparse and vectorized form, exploiting the inherent sparsity and parallelism of the discretized problem structure.

This paper presents a vectorized sparse second-order forward automatic differentiation methodology specifically designed for direct collocation methods in optimal control. The approach extends traditional expression graph structures to handle vectorized computations across mesh points while maintaining the theoretical rigor of forward-mode automatic differentiation. The resulting framework efficiently addresses the computational challenges of large-scale discretization while preserving analytical accuracy.

The methodology has been implemented in the open-source pockit package, providing a foundation for solving multi-phase optimal control problems. The implementation demonstrates the practical applicability of the theoretical framework and enables further research in optimal control methods.

\section*{Acknowledgments}
This work was supported by the National Natural Science Foundation of China (Grant No.12472355).

\printbibliography

\end{document}